\documentclass[aps,prl,twocolumn,groupedaddress]{revtex4}

\usepackage{graphicx}
\usepackage{dcolumn}
\usepackage{bm}

\begin{document}

\title{\large Quantum vortex tunneling in $YBa_2Cu_3O_{7-\delta}$ thin films}

\author{G. Koren, Y. Mor, A. Auerbach and E. Polturak }
\affiliation{Physics Department, Technion - Israel Institute of
Technology Haifa, 32000, ISRAEL}

\email{gkoren@physics.technion.ac.il}
\homepage{http://physics.technion.ac.il/~gkoren}

\date{\today}
\def\bfig {\begin{figure}[tbhp] \centering}
\def\efig {\end{figure}}

\normalsize \baselineskip=4mm  \vspace{15mm}

\begin{abstract}

Cuprate films offer a unique opportunity to observe vortex
tunneling effects, due  to their unusually low superfluid density
and short coherence length. Here, we measure the magnetoresistance
(\textit{MR}) due to vortex motion  of a long meander line of a
superconducting film made of underdoped $YBa_2Cu_3O_{7-\delta}$.
At low temperatures (\textit{T}), the \textit{MR} shows a
significant deviation from Arrhenius activation. The data is
consistent with two dimensional Variable Range Hopping (VRH) of
single vortices, i.e. $MR\propto exp[-(T_0/T)^{1/3}]$. The VRH
temperature scale $T_0$ depends on the vortex tunneling rates
between pinning sites. We discuss its magnitude with respect to
estimated parameters of the meander thin film.

\end{abstract}


\maketitle

Very soon after the discovery of the high temperature
superconductors, the resistive transition was observed to broaden
under magnetic fields, instead of shifting to lower temperatures
\cite{Palstra}. This was explained as due to thermally activated
vortex motion at high temperatures just below $T_c$, which gave
rise to an induced voltage across the superconductor and thereby
to a flux flow resistance $R_{ff}$. At any given magnetic field in
this regime, this resistance has the form of an Arrhenius law
$R_{ff} \propto exp[-(U_0/k_BT)]$, where $U_0$ is the activation
energy which was discussed by several authors
\cite{AndersonKim,Y&M,Tinkham}. Generally, flux flow and flux
creep are possible at high temperatures where the pinning is
relatively weak compared to the thermal energy. At low
temperatures, where the pinning is stronger and thermal activation
is much weaker, the dominant mechanism for flux motion is via
quantum tunneling. Measurements of magnetic relaxation and
transport by Stein \textit{et al.} \cite{Stein} have shown a
signature of quantum flux creep in
$Y_{1-x}Pr_xBa_2Cu_3O_{7-\delta}$ crystals. This behavior, where
the creep is temperature independent, was observed in a very
limited temperature range of about 2-4 K. Vortex tunneling in a 2D
superconductor at temperatures much lower than the transition
temperature $T_c$ was discussed theoretically by Fisher, Tokuyasu
and Young (FTY) \cite{Fisher}, and more recently by Auerbach,
Arovas and Gosh (AAG) \cite{AAG}. FTY studied quantum vortex
tunneling in 2D films at T close to zero near the superconductor
to insulator glass transition. They found that this tunneling
occurs via variable range hoping (VRH), but instead of the usual
1/3 power law in the exponent, at high fields they predicted a
different behavior where the exponent ranges between 2/3 and 4/5.
This resulted from taking into account vortex-vortex interactions.
At low fields ($H<<H_{c2}$), AAG calculated the tunneling rate of
a single vortex between two pinning sites and the resulting flux
tunneling resistivity. They found that this resistivity depends on
temperature as the well known VRH in 2D, namely $\rho\propto
exp[-(T_0/T)^{1/3}]$. In the present study we set up an experiment
to test this AAG prediction in a thin film of
$YBa_2Cu_3O_{7-\delta}$ (YBCO) patterned into a very long and
narrow meander line. We note that in a short microbridge of YBCO
of typically a few hundred $\mu m$ length, and under a magnetic
field of several Tesla, the voltage induced by moving vortices
becomes immeasurably small, and in practical terms, a critical
current develops already at about 10-20 K below $T_c$. This
prohibits measurements of $R_{ff}$ at low bias and low
temperatures. In contrast, in a much longer meander line, the
total DC component of the voltage generated by moving vortices is
large, and a de-facto resistive state persists down to very low
temperatures, thus enabling the vortex tunneling study.\\

A question still arises as for why quantum vortex tunneling has
not been unequivocally identified in low $T_c$ superconductors? We
believe that this is due to the fact that conventional BCS
superconductors are generally three dimensional and characterized
by a large coherence length and high superfluid density, all of
which greatly inhibit vortex tunneling. In contrast, underdoped
cuprate  films, such as investigated here, offer a unique system
of low superfluid density and quasi 2D superconductivity with very
short coherence lengths. According to a recent theoretical work
\cite{AAG}, in such  two dimensional "bosonic" superconductors,
vortex tunneling can be manifested in terms of deviation from
Arrhenius thermal activation of the magnetoresistance. Therefore,
observing the $(T_0/T)^{1/3}$ exponent would be an exciting
signature of the bosonic character of cuprate superconductivity as
well as a confirmation of short length-scale vortex tunneling
effects. Concerning previous experimental results, we note that
the phenomenon of vortex quantum creep has been investigated by
both magnetization relaxation and transport in three dimensional
cuprate crystals \cite{Stein}. Unfortunately, although a
temperature independent relaxation time and voltage drops have
been observed at low temperatures, these experiments did not
necessarily probe the low current regime where tunneling (rather
than above barrier classical diffusion) is expected to dominate.
The transport results explicitly show non-saturation at low
currents which indicates that no vortex tunneling occurred in the
low current, Ohmic regime. In our experiments, we study thin films
with low current bias. This ensures that deviation from Arrhenius
activation which we find is indeed related to vortex tunneling
and not to classical above-barrier motion.\\

In the present study, we used epitaxial \textit{c-axis} oriented
films of YBCO, with the magnetic field applied parallel to this
axis and perpendicular to the substrate wafer. The films were
prepared by 355 nm laser ablation deposition on (100) $SrTiO_3$
wafers of 10$\times10\, mm^2$ area. We investigated films of 50
and 100 nm thickness. In order to reduce their contact resistance,
some films were coated with a 20 nm thick gold layer deposited at
a high temperature (780 $^\circ$C) in oxygen ambient. This gold
overlayer consists of loosely connected ball-like grains as seen
by atomic force microscope images. Its contribution to the
resistance versus temperature of the bilayer was non-metallic and
very small. This was verified in a control experiment in which
another bilayer of 20 nm gold on 100 nm YBCO was prepared with the
YBCO doped to have a very low $T_c$. Thus, its resistance could be
measured down to very low temperatures. This bilayer was measured
as deposited, without patterning. After measuring the resistance,
the Au layer was removed using Ar ion milling, and the resistance
was measured again. A simple calculation of two parallel resistors
shows that the resistivity of the gold layer below 10 K was above
5 $m\Omega\,cm$. This value is much higher than the resistivity of
the 60 K YBCO phase which is about 0.1 $m\Omega\,cm$ at 100 K.
Considering also the 1:5 thickness ratio between the gold and YBCO
layers used, we find that the presence of the gold layer changed
the resistance of the meander line by about 0.25$\%$, which is
negligible. Nevertheless, the aim of obtaining a low contact
resistance with the YBCO film where the contact is made normal to
the surface was certainly achieved. We note that the \textit{MR}
results measured with and without the gold layer were basically
the same. However, those obtained with the gold overcoating were
less noisy, and therefore we chose to present them here.\\

The YBCO films and Au/YBCO bilayers were annealed \textit{in-situ}
in a controlled oxygen ambient to have a transition temperature of
50-60 K, in order to avoid apparent critical currents at low
temperatures. We note that the presence of a critical current
interferes with the resistance studies at low bias where the
current should be proportional to the voltage. The films were
patterned by deep UV lithography using a PMMA resist and Ar ion
beam milling, into a 4 m long meander line. This meander line had
14 $\mu m$ lines-width with 4 $\mu m$ lines-spacing on $8\times
9\,mm^2$ area of the wafer, while the remaining area was used for
the four $1.5\times 1.5\,mm^2$ contacts. In the patterning
process, we limited the development time of the photo-resist in
order to keep it continuous, but this left several shorts,
effectively shortening the meander line. In addition, we found
that in the patterning process a few defects were formed in the
long meander line which we bridged with small silver paste dots of
$\sim 0.5\,mm$ diameter. The gold overcoating layer was also very
helpful in improving the contacts to these bridging dots. This
bridging procedure and the shorts mentioned before led to a
reduced effective length of our meander lines of about 1 m. This
length was estimated from the measured resistance, the resistivity
and the cross section area of the bare YBCO meander lines without
the gold coating. For the transport measurements we used the
standard four contacts technique in a dc mode, and cooling was
done in a liquid helium cryostat with a base
temperature of 2 K.\\

\begin{figure}
\hspace{-5mm}\includegraphics[height=7cm,width=9cm]{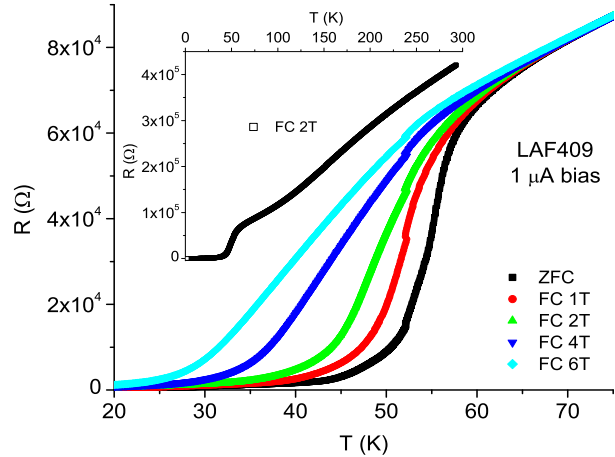}
\caption{\label{fig:epsart}(Color online) Resistance versus
temperature of the YBCO meander line (of $\sim$1m length and
$100\,nm \times 14\,\mu m$ cross section area) under various
cooling conditions with and without a magnetic field. The
resistance of the meander line over a wider temperature range
under 2 T field cooling is shown in the inset.}
\end{figure}

\begin{figure}
\includegraphics[height=7cm,width=9cm]{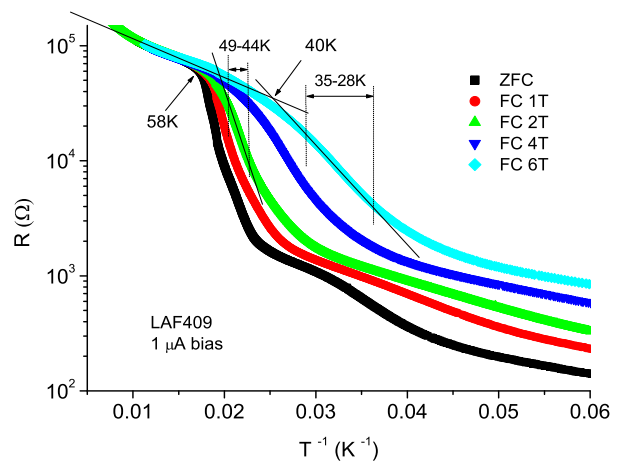}
\caption{\label{fig:epsart}(Color online) The resistance data of
Fig. 1 plotted on a log scale versus inverse temperature. The
standard vortex activation regions where $R\propto exp(-U_0/kT)$
are shown for fields of 2 T and 6 T.}
\end{figure}

The resistance versus temperature results of the meander line are
shown in Fig. 1 for zero field cooling (ZFC) and  under  magnetic
fields of 1, 2, 4, and 6 T applied normal to the wafer. One can
see the typical behavior characteristic of underdoped YBCO in the
2-300 K temperature range in the inset to this figure. In the main
panel, the superconducting transition temperature range is shown
in more detail. The well known thermally activated broadening of
the transition under increasing magnetic field is clearly
observed, and in addition a significant resistive tail is seen
already at zero field. To see the thermal activation functional
dependence, and the data at low temperatures, the data of Fig. 1
was reploted in Fig. 2 on a log scale versus inverse temperature.
One can see that the regimes where thermal activation holds  are
small, about 5 K at 2 T and 7 K at 6 T. The onset of the
superconducting transition under ZFC is $T_c^{onset}=58\,K$, while
that of the 6 T field cooled curve is shifted down to about 40 K.
This is qualitatively similar to the results of Palstra \textit{et
al.}, but with about twice the measured shift of the 90 K phase of
YBCO single crystals \cite{Palstra1990}. Also seen in Fig. 2 is a
knee in the ZFC curve at 42-34 K just below the main transition.
This is a clear signature of weak links, which is very similar to
the results on grain boundary junctions \cite{Gross}. These weak
links can be due to defects in the STO substrate which are copied
into the epitaxial film, presence of grains of the minority
($\sim$3\%) \textit{a-axis} oriented phase, lithography induced
defects and so on. It is therefore clear that any possible vortex
tunneling unmasked by other effects could be observed only at
temperatures
below about 20-30 K.\\

\begin{figure}
\hspace{-7mm}\includegraphics[height=7cm,width=9cm]{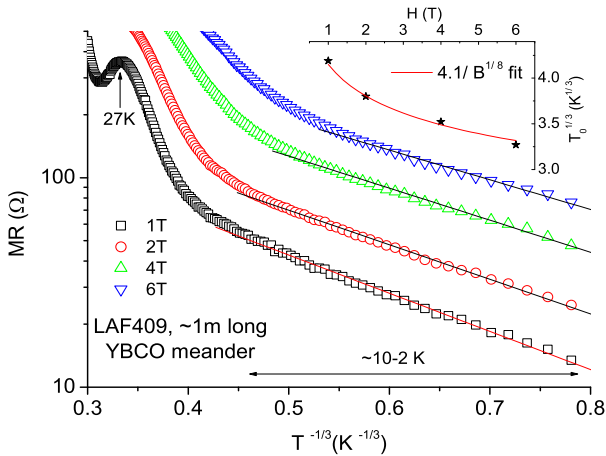}
\caption{\label{fig:epsart}(Color online) The magnetoresistance
$MR=R(H)-R(0)$ derived from the data of Fig. 1, plotted on a log
scale as a function of $T^{-1/3}$. The straight lines are linear
fits to the data in the relevant low temperature regimes. In the
inset the slopes of these straight lines (on a ln scale)
$T_0^{1/3}$ are plotted versus field, together with a fit to
$4.1/B^{1/8}$. }
\end{figure}

\begin{figure}
\hspace{-5mm}\includegraphics[height=7cm,width=9cm]{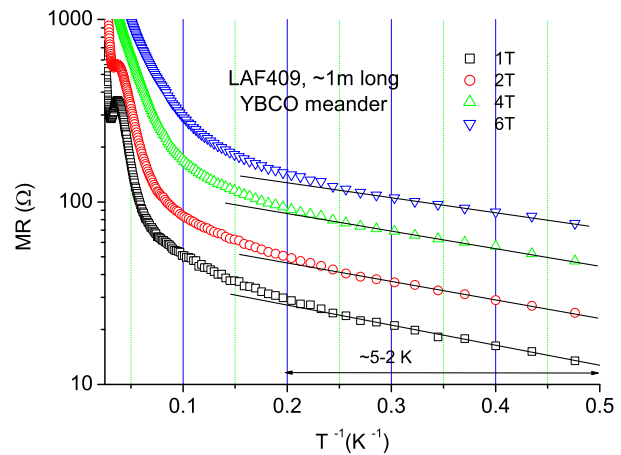}
\caption{\label{fig:epsart}(Color online) The data of Fig. 3
plotted versus inverse temperature to compare possible standard
vortex activation to vortex tunneling by VRH as in Fig. 3. The
straight lines are guides to the eye. }
\end{figure}

Next we study the temperature dependence of the magnetoresistance
of the meander line. We use the usual definition of $MR=R(H)-R(0)$
where $R(H)$ and $R(0)$ are the resistances under a magnetic field
H and at zero field, respectively. Using the magnetoresistance
rather than the resistance under field itself has the advantage of
eliminating all extrinsic effects such as those contributed by
weak links and series resistors of the repairing silver paste
dots. We can safely assume that these contributions to the
magnetoresistance are small since the relevant defects occupy a
very small fraction of the area of the meander line, and their
resistance has a much weaker dependence on the perpendicular
magnetic field. In Fig. 3 the $MR$ is plotted versus $T^{-1/3}$
where a linear dependence should indicate VRH in the YBCO planes,
while in Fig. 4 the $MR$ is plotted versus $1/T$ to test for a
possible Arrhenius activation process. As one can see a linear
behavior is found in both figures at low temperatures, but the VRH
range of Fig. 3 extends over 2-10 K which is about three times as
large as the activation range of 2-5 K as seen in Fig. 4. Although
this temperature range is relatively small, the fact that the
$log(MR) \propto T^{-1/3}$ behavior exists in a significantly
larger temperature range suggests that we actually observe VRH as
predicted by AAG \cite{AAG}. Their expression for the
magneto-resistivity in a 2D superconductor is:

\begin{equation}
MR(B,T)=\big{(}\frac{h}{2e}\big{)}^2\gamma_0[n_v(B)]exp[-(T_0/T)^{1/3})]
\end{equation}

\noindent where $\gamma_0$ is the vortex conductivity which
depends on the vortex density $n_v(B)$, and $T_0$ is given by:

\begin{equation}
T_0=K\delta \overline{V}\big{(}\frac{\pi n_s}{n_{pin}}\big{)}^2
\end{equation}

\noindent where $K$ is a dimensionless factor of order unity,
$\delta \overline{V}$ is the average variation of the pinning
energy, $n_s$ is the pairs density and $n_{pin}$ is the pinning
sites density. It should be noted that this theory strictly
applies only to the 2D case, while the actual film has a quasi 3D
nature with pancake vortices which have very weak Josephson
coupling in the c direction. The vortices in the film, form
straight lines which minimize the total pinning potentials of the
layers. The minimal instanton action (wavefunction overlap) for
the tunneling events involves motion of an  individual vortex
which has a pinning site in its close neighborhood. After the
tunneling event, the vortex lines up to its new position via
classical (above barrier) relaxation. This satisfies the
ingredients of the variable range hopping model, where the mean
tunneling distance in a thin film is of the order of
$l_{2D}/\sqrt{ N_{layers}}$, where $l_{2D}$ is the mean distance
between uncorrelated pinning sites in each 2D layer, and
$N_{layers}$ is the number of layers in the film. Therefore, for
the quasi 3D case of a thin film, Eq. (2) has to be modified by
replacing $n_{pin}$ by $n_{pin}\times N_{layers}$ which yields:

\begin{equation}
T_0(film)=K\delta \overline{V}\big{(}\frac{\pi
n_s}{n_{pin}N_{layers}}\big{)}^2.
\end{equation}

One can see that for any given field the results of Fig. 3 at low
temperatures are in good agreement with Eq. (1). Although the FTY
theory \cite{Fisher} is applicable only at high fields, we also
tested their prediction of $\rho \propto exp[-(T_0/T)^{2/3})]$,
but this dependence fit our $MR$ data only in the narrow regime of
2-5 K. In the inset of Fig. 3, the $T_0^{1/3}$ coefficient of Eq.
(1) is plotted versus field. We see that this coefficient is not
exactly  a constant as assumed for low fields by the theory, but
has a small decreasing contribution of about 30\% with increasing
magnetic field. This behavior can be explained by the effect of
decreasing barrier height for vortex tunneling when the
vortex-vortex interaction is increased with increasing field.\\

Next we make a consistency check of our data at H=2 T with Eq.
(3). From the data of Fig. 2 just below $T_c$ we find that the
pinning energy of our films is $U_0=V\approx 550\,K$. For $\delta
\overline{V}$ of about 10\% of $V$ one gets $\delta
\overline{V}\approx 55\,K$. The measured $T_0$ from the inset of
Fig. 3 at 2 T is $3.8^3\approx 55\,K$. Thus, for our data Eq. (3)
implies that $\pi n_s \approx n_{pin}N_{layers}$. For pairs, $n_s$
equals half the doping $p$ per copper in the $CuO_2$ plane. In our
60 K YBCO phase, $p\approx 0.12$ \cite{Segawa} which yields
$n_s\approx 0.06$ per copper. Thus $n_{pin}\approx \pi 0.06/170$
per copper in a single $CuO_2$ plane, where $N_{layers}\approx
170$ is the number of $CuO_2$ planes in the 100 nm thick YBCO
film. This implies that the average distance  between pinning
sites in a single $CuO_2$ plane is $l_{2D}\approx 3.9/\sqrt{\pi
0.06/170}\approx 117\,\rm\AA$ which is very reasonable. (the
3.9$\rm \,\AA$ here is the in-plane lattice constant of YBCO).\\

\begin{figure}
\hspace{-5mm}\includegraphics[height=7cm,width=9cm]{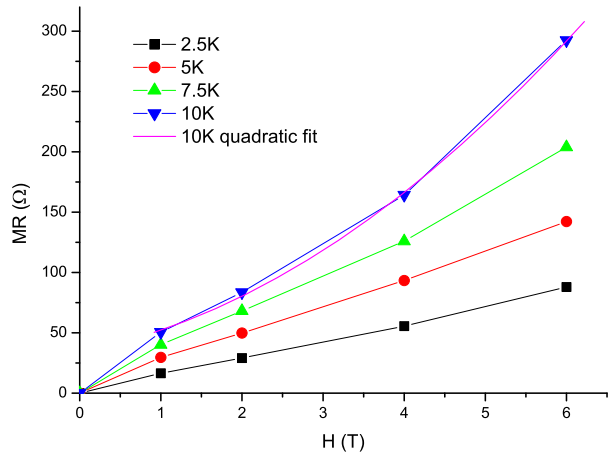}
\caption{\label{fig:epsart}(Color online) Magnetoresistance as a
function of magnetic field at a few temperatures. The straight
lines are just connecting the data points, while the parabolic
curve is a quadratic fit to the data at 10 K for fields of 1-6 T.}
\end{figure}

In Fig. 5 we plot the measured magnetoresistance $MR$ versus
magnetic field for a few temperatures. One can see that at the low
temperatures of 2.5 and 5 K the $MR$ is almost linear in field. In
contrast, at higher temperatures the dependence on the field is
more complex, as seen from the data at 7.5 and 10 K. The linear
$MR$ dependence on field at a constant temperature can be
attributed to a constant terminal vortex velocity in the YBCO
film. This would give rise to an induced voltage across the
meander line which increases linearly with field. The  behavior of
the $MR$ at higher temperatures and higher fields can originate in
the weak links which are distributed along the meander line. Under
these conditions, the weak links become normal, and this gives
rise to an $MR$ which reflects the distribution of the strength of
the weak links which can be quite complex.\\

Finally, we point out that the main difference between the present
study and the work on the vortex glass dynamics \cite{Fisher}, is
that here we probed the low field (low vortex density) regime. The
single vortex tunneling theory \cite{AAG} predicts a
magnetoresistance linear in H as observed in Fig. 5, and ignores
vortex interaction effects which are of higher order in the
magnetic field. Experimentally, this is justified since at the
highest field used of 6 T, the distance between vortices is of
about 20 nm, while the distance between pinning sites in the whole
film is only $117/\sqrt{170}\approx 0.9$ nm. Therefore, collective
tunneling effects might well be relevant only at high fields and
temperatures. Since our experiment probes the low field regime,
(much lower than $H_{c2}$), we expect single vortex variable range
hopping to dominate our results.\\

In conclusion, vortex variable range hopping in two dimensions is
consistent with our experimental magnetoresistance results in a
long meander line of underdoped YBCO thin films at low
temperatures.\\

{\em Acknowledgments:}  AA acknowledges useful discussions with
Steve Kivelson and Bert Halperin. This research was supported in
part by the Israel Science Foundation (grant \# 1564/04), the
US-Israel Binational Science Foundation, the Heinrich Hertz
Minerva Center for HTSC, the Karl Stoll Chair in advanced
materials, and by the Fund for the Promotion of Research at the
Technion.\\

\bibliography{AndDepBib.bib}

\bibliography{apssamp}

\begin{thebibliography}{99}
\label{Bib}

\bibitem{Palstra} T. T. M. Palstra, B. Batlogg, L. F. Schneemeyer
and J. V. Waszczak, Phys. Rev. Lett. \textbf{61}, 1662 (1988).
\bibitem{AndersonKim} P. W. Anderson and Y. B. Kim, Rev. Mod.
Phys. \textbf{36},  39 (1964).
\bibitem{Y&M} Y. Yeshurun and A. P. Malozemoff, Phys. Rev. Lett.
\textbf{60}, 2202 (1988).
\bibitem{Tinkham} M. Tinkham, Phys. Rev. Lett. 61, 58 (1988).
\bibitem{Stein} T. Stein, G. A. Levin, C. C. Almasan, D. A.Gajewski
and M.B.Maple, Phys. Rev. Lett. \textbf{82}, 2955 (1999).
\bibitem{Fisher} M. P. A. Fisher, T. A. Tokuyasu and A. P. Young,
Phys. Rev. Lett. \textbf{66}, 2931 (1991).
\bibitem{AAG} A. Auerbach, D. P. Arovas and S. Ghosh, Phys. Rev. B
\textbf{74}, 064511 (2006).
\bibitem{Palstra1990} T. T. M. Palstra, B. Batlogg, R. B. Van Dover
L. F. Schneemeyer and J. V. Waszczak, Phys. Rev. B \textbf{41},
6621 (1990).
\bibitem{Gross} R. Gross, P. Chaudhari, D. Dimos, A. Gupta and G.
Koren, Phys. Rev. Lett. \textbf{64}, 228 (1990).
\bibitem{Segawa} Y. Ando and K. Segawa, Phys. Rev. Lett.
\textbf{88}, 167005 (2002).






\end{thebibliography}

\end{document}